\documentclass[peerreview]{IEEEtran}
\IEEEoverridecommandlockouts
\usepackage{amsmath,amssymb,amsfonts,amsthm}
\DeclareMathAlphabet\mathbfcal{OMS}{cmsy}{b}{n}

\usepackage[dvipsnames]{xcolor}

\usepackage{enumerate}
\usepackage{xfrac}
\usepackage{graphicx}
\usepackage{tikz}
\usepackage{makecell}
\usepackage{textcomp}
\usepackage{textgreek}
\usepackage{svg}
\usepackage[dvipsnames]{xcolor}
\usepackage[colorlinks=true,linkcolor=black,citecolor=black,urlcolor=black]{hyperref}
\def\BibTeX{{\rm B\kern-.05em{\sc i\kern-.025em b}\kern-.08em
    T\kern-.1667em\lower.7ex\hbox{E}\kern-.125emX}}

\usepackage{url}
\usepackage{float}
\usepackage{multirow}
\usepackage{color}
\usepackage[nolist,printonlyused]{acronym}    
\usepackage{comment}
\usepackage{bm}
\usepackage{comment}
\usepackage{algorithm}
\usepackage{algorithmicx}
\usepackage{algpseudocode}
\usepackage{pgf}
\usepackage{pgfplots}
\pgfplotsset{compat=newest}
\usepackage{import}
\usepackage{bbm}
\usepackage{tikz}
\usetikzlibrary{spy}
\usetikzlibrary{arrows,decorations.pathreplacing,patterns}
\usepackage[sorting=none,style=ieee, minbibnames=4, maxbibnames=4, url=false, doi=false]{biblatex} 

\AtBeginBibliography{\footnotesize} 

\usepackage{csquotes}

\definecolor{airforceblue}{rgb}{0.36, 0.54, 0.66}

\definecolor{eblue}{HTML}{0082F0}
\definecolor{egreen}{HTML}{0FC373}
\definecolor{epurple}{HTML}{AF78D2}
\definecolor{eyellow}{HTML}{FAD22D}
\definecolor{eorange}{HTML}{FF8C0A}
\definecolor{ered}{HTML}{FF3232}

\definecolor{eblue3}{HTML}{B2D8F9}
\definecolor{egreen3}{HTML}{BDEDD6}
\definecolor{epurple3}{HTML}{E8D6F2}
\definecolor{eyellow3}{HTML}{FEF1C6}
\definecolor{eorange3}{HTML}{FFDBB5}
\definecolor{ered3}{HTML}{FEC3C2}

\begin{document}

\title{Advancing Network Digital Twin Frameworks \\ for Generating Realistic Datasets \\ 

\thanks{\copyright 2026 IEEE. Personal use of this material is permitted. Permission from IEEE must be obtained for all other uses, in any current or future media, including reprinting/republishing this material for advertising or promotional purposes, creating new collective works, for resale or redistribution to servers or lists, or reuse of any copyrighted component of this work in other works.

Oscar Stenhammar, Gabor Fodor, and Carlo Fischione are with the School of Electrical Engineering and Computer Science, KTH Royal Institute of Technology, Stockholm, Sweden (e-mail: ostenh@kth.se, gaborf@kth.se, carlofi@kth.se).
Oscar Stenhammar and Gabor Fodor are also with Ericsson Research, Sweden (e-mail: oscar.stenhammar@ericsson.com, gabor.fodor@ericsson.com). Sundeep Rangan is with the Electrical \& Computer Engineering Department, NYU Tandon (srangan@nyu.edu)

}
}

\author{
Oscar Stenhammar,
Sundeep Rangan,
Gábor Fodor,
Carlo Fischione
}

\begin{acronym}
  \acro{3GPP}{3$^\text{rd}$~Generation Partnership Project}
  \acro{4G}{4$^\text{th}$~Generation}
  \acro{5G}{5$^\text{th}$~generation}
  \acro{6G}{6$^\text{th}$~generation}
  \acro{AI}{artificial intelligence}
  \acro{AC}{admission control}
  \acro{ANN}{artificial neural network}
  \acro{AR}{autoregressive}
  \acro{ARMA}{autoregressive moving average}
  \acro{ARIMA}{auto regressive integrated moving average}
 \acro{AWGN}{additive white Gaussian noise}
  \acro{BPNN}{back propagation neural network}
  \acro{BER}{bit error rate}
  \acro{BPSK}{binary phase-shift keying}
  \acro{BS}{base station}
  \acro{CDF}{cumulative distribution function}
  \acro{CNN}{convolutional neural network}
  \acro{CSI}{channel state information}
  \acro{CSIR}{channel state information at the receiver}
  \acro{CSIT}{channel state information at the transmitter}
  \acro{C-ITS}{cooperative ITS}
  \acro{CUE}{cellular user equipment}
  \acro{DL}{downlink}
  \acro{DNN}{deep neural network}
  \acro{D-MIMO}{distributed multiple input multiple output}
  \acro{eNb}{e Node b}
  \acro{FDD}{frequency division duplexing}
  \acro{FL}{federated learning}
  \acro{GRU}{gated recurrent unit}
  \acro{IID}{independent and identically distributed}
  \acro{IoT}{Internet of Things}
  \acro{ITS}{intelligent transportation systems}
  \acro{KDE}{kernel density estimation}
  \acro{KF}{Kalman filter}
  \acro{KPI}{key performance indicator}
  \acro{LoS}{line of sight}
  \acro{LS}{least squares}
  \acro{LSTM}{long short-term memory}
  \acro{LTE}{long term evolution}
  \acro{MAC}{medium access control}
  \acro{MAE}{mean absolute error}
  \acro{MAPE}{mean absolute percentage error}
  \acro{mMIMO}{massive multiple input multiple output}
  \acro{MIMO}{multiple input multiple output}
  \acro{ML}{machine learning}
  \acro{MLP}{multilayer perceptron}
  \acro{MMSE}{minimum mean squared error}
  \acro{MSE}{mean squared error}
  \acro{MU-MIMO}{multiuser multiple input multiple output}
  \acro{NDT}{network digital twin}
  \acro{NLOS}{non-line of sight}
  \acro{NMSE}{normalized mean square error}
  \acro{NN}{neural network}
  \acro{OFDMA}{orthogonal frequency division multiple access}
  \acro{OFDM}{orthogonal frequency division multiplexing}
  \acro{pQoS}{predictive QoS}
  \acro{PRB}{physical resource block}
  \acro{QoS}{quality of service}
  \acro{QoE}{quality of experience}
  \acro{RAN}{Radio Access Network}
  \acro{RAT}{radio access technology} 
  \acro{ReLU}{rectified linear unit}
  \acro{RL}{reinforcement learning}
  \acro{RF}{radio frequency}
  \acro{RSCP}{Received Signal Code Power}
  \acro{RSRP}{Reference Signal Receive Power}
  \acro{RSRQ}{Reference Signal Receive Quality}
  \acro{RSSI}{Received Signal Strength Indicator}
  \acro{SLA}{service level agreement}
  \acro{SNR}{signal-to-noise ratio}
  \acro{SVD}{singular value decomposition}
  \acro{SVT}{singular value thresholding}
  \acro{TDD}{time division duplexing}
  \acro{TDL}{tapped delay line}
  \acro{UE}{user equipment}
  \acro{UL}{uplink}
  \acro{V2V}{vehicle-to-vehicle}
  \acro{V2X}{vehicle-to-everything}
  \acro{ZF}{zero-forcing}
  \acro{ZMCSCG}{zero mean circularly symmetric complex Gaussian}
\end{acronym}


\maketitle

\begin{abstract} 

The integration of accurate and reproducible wireless network simulations is a key enabler for research on open, virtualized, and intelligent communication systems. Network Digital Twins (NDTs) provide a scalable alternative to costly and time-consuming measurement campaigns, while enabling controlled experimentation and data generation for data-driven network design. In this paper, we present an open and user-friendly NDT framework that integrates controllable vehicular mobility with the site-specific ray tracer Sionna and the discrete-event ns-3 network simulator, enabling virtualized end-to-end modeling of wireless networks across the radio, network, and application layers. The proposed framework is particularly well-suited for dynamic vehicular networks and urban deployments, supporting realistic mobility, traffic dynamics, and the extraction of cross-layer metrics. To promote open-source initiatives, we release both the NDT implementation and a representative dataset generated from realistic vehicular and urban scenarios. The framework and dataset facilitate reproducible experimentation and benchmarking of machine learning–based quality-of-service prediction, network optimization, and intelligent network management algorithms, lowering the entry barrier for research on virtual and open wireless network services.


\end{abstract}



\section{Introduction}


\IEEEPARstart{T}{he} advent of sixth-generation (6G) wireless communication technologies is expected to enable highly flexible, virtualized, and intelligent communication systems~\cite{10745245}. 6G is envisioned to support open and software-driven network architectures capable of adapting to diverse vertical applications, including \ac{ITS} and connected and cooperative automated mobility (CCAM)~\cite{11126933}. Such applications impose stringent requirements on reliability and latency, where communication failures or excessive delays may lead to severe safety and operational consequences. Consequently, future network operators must provision communication services that can sustain reliable and low-latency performance under pronounced spatiotemporal variability in traffic demand, mobility, and radio propagation conditions~\cite{9503121}.

To address these challenges, \acp{NDT} are emerging as a key enabler for open and virtualized network operation. An \ac{NDT} constitutes a fine-grained, software-based replica of a communication network that mirrors the behavior of its physical counterpart and enables experimentation without disrupting live services~\cite{9999174}. By supporting reproducible, controllable, and scalable experimentation, NDTs facilitate data-driven network management tasks such as performance prediction, optimization, and service assurance. As such, NDTs provide a natural foundation for developing and validating \ac{ML}–based solutions in complex and dynamic wireless environments.

Achieving realism in NDTs requires several components to be fulfilled. Wireless propagation and mobility models must capture the dynamics of real environments to reflect time-varying interference and fading caused by mobility patterns. The NDT frameworks must have the ability to scale to the complexity of large and heterogeneous networks while maintaining sufficient granularity to represent distributed devices. Modeling the application layer and the behavior of end users is essential to reproduce realistic traffic patterns and capture the nature of the load experienced in operational networks. If the NDT is installed in a live network, leveraging measurements from real-time systems together is increasingly crucial to calibrate and update the NDT. The combination of these important elements, among others, enables NDTs to evolve beyond static simulations into continuously synchronized systems that can help networks adapt to dynamic network conditions and remain aligned with real-world performance~\cite{11077799}.

Recent work has emphasized the need for realistic, controllable, and scalable \acp{NDT} to support reproducible research for vehicular and urban wireless systems~\cite{11077799,10945722, 10879345}. This has sparked collective contributions to creating more realistic and accessible NDT platforms. 
The ms-van3t framework from the work in~\cite{RAVIGLIONE202470} provides an open-source, standardization-compliant simulation environment that integrates the vehicular mobility simulator SUMO and network simulator ns-3~\cite{Riley2010} to support \ac{V2X} simulations. 
The work in~\cite{10578207} introduces OScar, an open and lightweight \ac{C-ITS} stack designed for affordable vehicular field tests.
At the infrastructure scale, the Colosseum simulator~\cite{10643670} demonstrates how large radio frequency emulators and software protocol stacks can realize Open \ac{RAN} digital twins for end-to-end experimentation. 

Traditional discrete‐event network simulators such as ns‑3~\cite{Riley2010} are well‐suited for modeling protocol stacks and network traffic flows. However, they often rely on simplified channel models that cannot capture fine‐grained spatial, temporal, and environmental variations. Ray-tracing-based frameworks such as Sionna RT~\cite{hoydis2022sionna} can produce detailed channel realizations, but lack the integration of full protocol stack behaviour.  
The work in~\cite{11152946} integrates ns-3 with Sionna RT to create the first full-stack open-source NDT, which enables deterministic ray-traced channel modeling.

\begin{figure}[t]
    \centering
    \includegraphics[trim={0 0 0 1cm}, clip,width=0.55\linewidth]{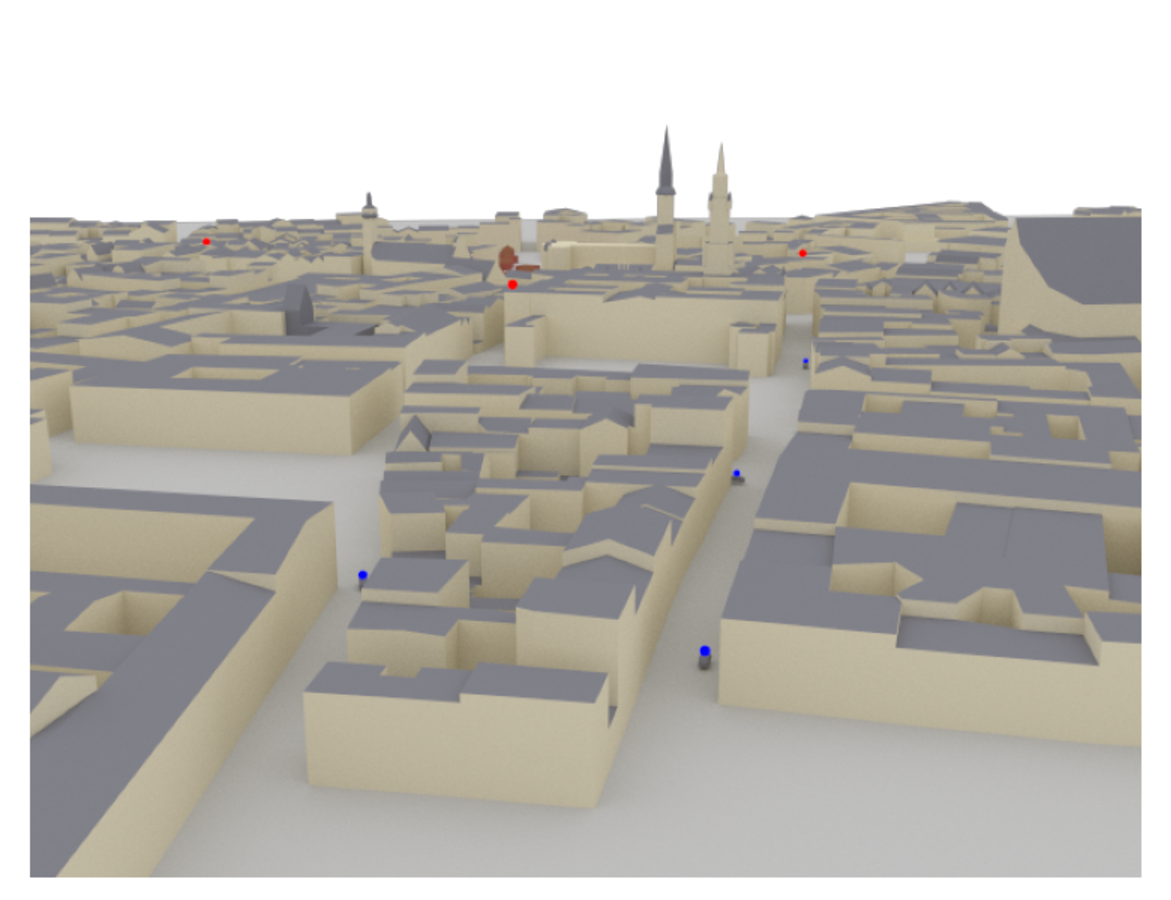}
    \caption{This figure illustrates the considered scene in Sionna. Base station antennas are marked by red spheres, and blue spheres indicate the placement of the antennas of the vehicles.}
    \label{fig:scene}
\end{figure}

The study in~\cite{10213404} shows that real network data is subject to concept drift, meaning that traffic patterns, channel conditions, and user behavior can change significantly over time. This highlights the need for an NDT capable of generating heterogeneous and evolving network scenarios so that \ac{ML} models can be trained and validated under realistic, non-stationary conditions.

Realistic NDTs have the ability to empower wireless network research and evaluation of \ac{ML} algorithms. However, there is a gap for open and accessible NDT frameworks that unify ray-tracing, user mobility, and full network stack simulations for scalable ML-oriented research. We extend prior work in~\cite{11152946}, which integrates Sionna RT with ns-3 to support full-stack network data flows over channels computed by ray-tracing, by developing a framework that incorporates user mobility and varying network load into the model. This framework significantly enhances the simulation model by supporting a high number of controllable mobile devices and vehicles. Although SUMO generates realistic vehicle mobility, our framework allows for greater controllability. The network load is implemented to follow a realistic pattern over the hours of the day, while maintaining a degree of stochasticity to reflect authentic network behaviors. This user-friendly NDT provides a rigorous baseline implementation, but can be easily configured to accommodate the desired experimental setup. The simulation framework supports dynamic vehicular mobility with a heterogeneous environment and trajectory selection in real urban grids. A comprehensive logging system is integrated, enabling data sources from the entire network stack. This produces an accessible and reproducible environment towards an NDT ideal for research into dynamic vehicular networks, adaptive communication strategies, and \ac{ML}‐based algorithms. To demonstrate its utility, we provide an open‐source repository providing the NDT tool\footnote{https://github.com/osst3224/ns3-rt-mobility} that includes extensive examples. To make this tool even more accessible, a representative dataset is published~\cite{81qg-r354-25}, generated from network load and vehicular mobility reflecting reality. The resulting dataset includes spatial, temporal, and network data, and serves as a rich foundation for \ac{ML} tasks such as predictive \ac{QoS}, anomaly detection, and adaptive resource management.


The remainder of this paper is organized as follows. The simulation setup is described in Section~\ref{sec:sim}. A data analysis is conducted based on the generated data, and a simple \ac{QoS} prediction example is provided, in Section~\ref{sec:data}. We present a discussion of the presented work and potential studies with the generated data for future work in Section~\ref{sec:dis}. Finally, we conclude our work with a conclusion in Section~\ref{sec:con}.







\begin{figure}[t]
    \centering
    \includegraphics[width=0.75\linewidth]{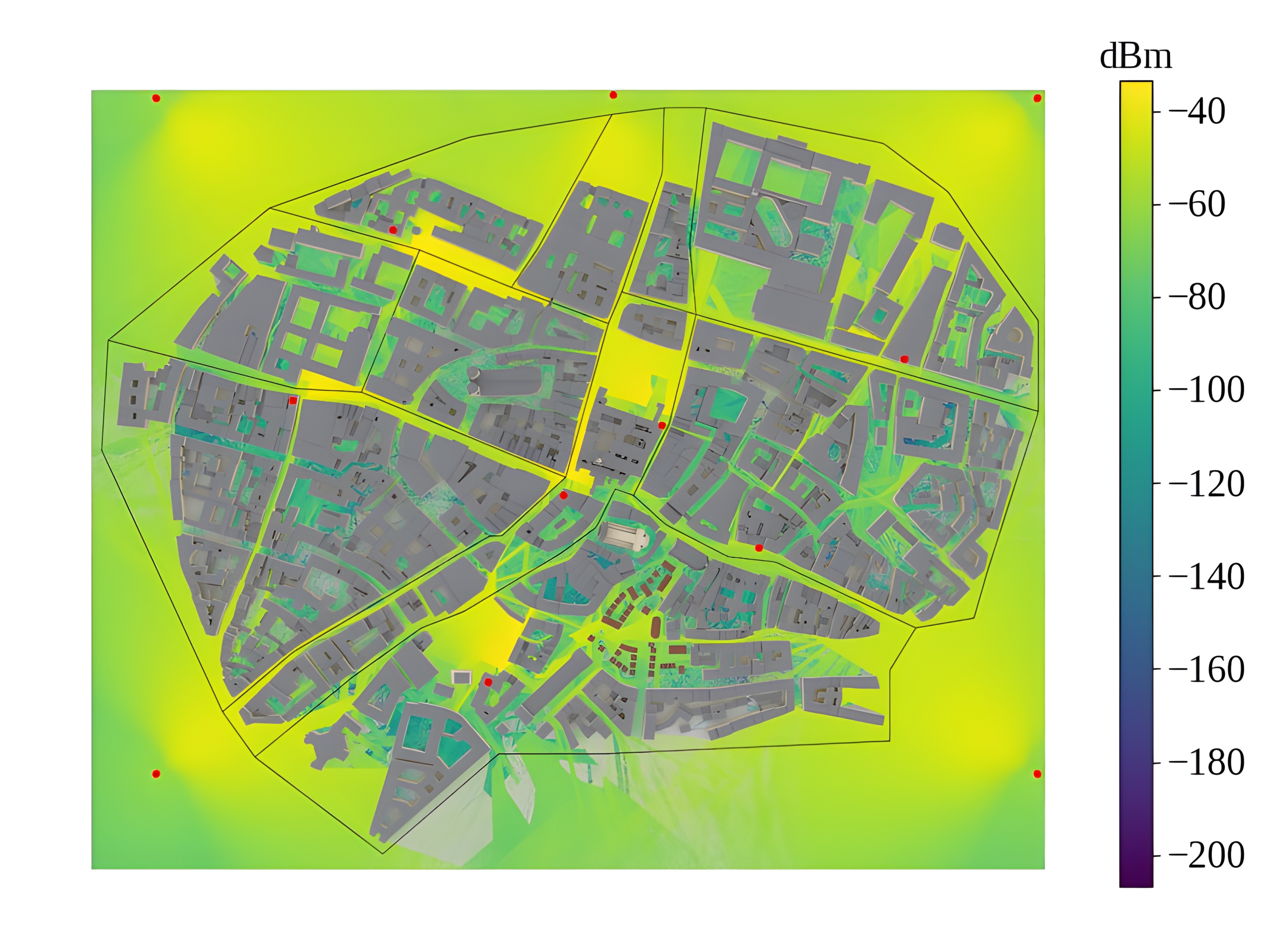}
    \caption{A map of the signal strength in the virtual scenario. Base station antennas are marked by red spheres. The black lines indicate the streets along which vehicles can travel.}
    \label{fig:heatmap}
\end{figure}

\section{Simulation Setup}
\label{sec:sim}
To obtain a realistic NDT, we simulate a wireless network by combining the ns-3 discrete-event simulator with the Sionna link-level ray-tracer based on the work in~\cite{11152946}. Ray-traced propagation data is generated in the predefined 500 × 500 m Munich scenario in Sionna and imported directly into ns-3, replacing standard stochastic channel models and enabling site-specific modeling of path loss, shadowing, and multipath effects. A fraction of the scene is visualized in Fig.~\ref{fig:scene}.

Vehicular mobility of the users is modeled by simulating connected cars that traverse the road network of the Munich scene. The number of active vehicles is an input parameter. We set probabilities that define crowded and less-populated regions of the map and spawn vehicles accordingly. We introduce a random number generator that assigns each vehicle a trajectory, with a higher likelihood of selecting routes through crowded areas and a lower likelihood of entering sparsely populated ones. Within each region, we apply area-specific speed limits between 30 km/h to 100 km/h to reflect normal traffic conditions. This approach allows us to generate heterogeneous and realistic mobility patterns throughout the simulated urban environment.

To emulate realistic network dynamics, we model background traffic in the network using a 24-hour diurnal load profile; low during early morning hours, peaking during rush hours, and moderate throughout the day.
We place 12 base stations with one antenna unit each, and set inter-site distances to ensure coverage of the simulated area. The signal strength is plotted as a heatmap in the considered scenario in Fig.~\ref{fig:heatmap}. We set the network to operate at 3.5 GHz with 20 MHz bandwidth and a maximum transmit power of 30 dBm.

Within this environment, ns-3 instantiates the network stack and applies the time-varying traffic load, while Sionna calculates the link-level radio conditions. The user mobility that we implement is managed in the ns-3 part of the code. To collect multi-layer data, we incorporate a FlowMonitor from ns3 with a sampling period of one second. Logged features include flow identifiers, serving cell load, UE position, velocity, and direction, as well as packet-level statistics such as transmitted and received packet error rate, packet sizes, end-to-end latency, throughput, and jitter. Radio-level measurements capture SINR, RSRP, and \ac{LoS} status. By running repeated simulations across varying traffic and mobility conditions, we generate a large-scale dataset~\cite{81qg-r354-25} suitable for training and evaluating \ac{ML} models for applications that require a heterogeneous dataset. 


\section{Data Analysis}
\label{sec:data}
\subsection{Simulation Setup}
\begin{figure}[t]
    \centering
    \includegraphics[width=0.65\linewidth]{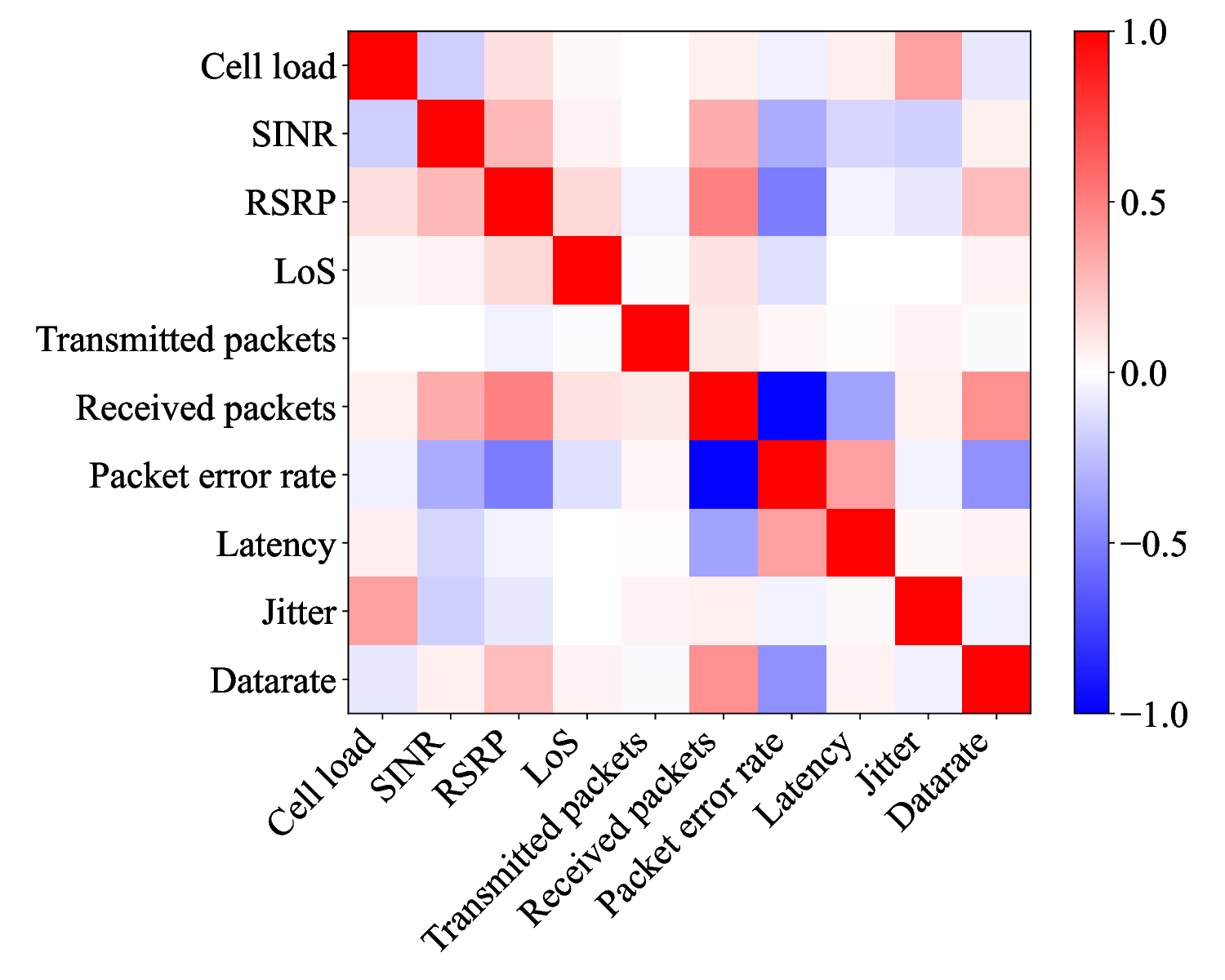}
    \caption{Correlation matrix of network-related features, illustrating pairwise Pearson correlation coefficients between traffic load, radio conditions, packet-level statistics, and end-to-end performance metrics. The color scale ranges from -1, representing strong negative correlation, to 1, representing strong positive correlation.}
    \label{fig:corr}
\end{figure}

We provide an analysis of the generated dataset that we provide in~\cite{81qg-r354-25}, focusing on diversity among the different cells and the interplay between radio and traffic data. The correlation matrix in Fig.~\ref{fig:corr} summarizes the linear dependencies among the key QoS metrics in the dataset. As expected, the channel-quality metrics SINR and RSRP are positively correlated. Both SINR and RSRP show a moderate negative correlation with latency and packet error rate, reflecting their role as primary drivers of link reliability. \ac{LoS} conditions also correlate positively with SINR and RSRP, indicating that \ac{LoS} propagation is associated with stronger received power and improved signal quality. All simulation parameters used for this simulation are included in the provided GitHub repository.

On the traffic side, packet error rate shows a strong negative correlation with received packets and a positive correlation with latency, consistent with congestion- or interference-driven performance degradation. The datarate correlates positively with good radio conditions and negatively with packet error rate and cell load. Overall, the matrix reflects the expected interplay between radio conditions, traffic load, and the resulting network performance.

\begin{figure}
    \centering
    \includegraphics[trim={0 0 0 0.4cm}, clip, width=0.65\linewidth]{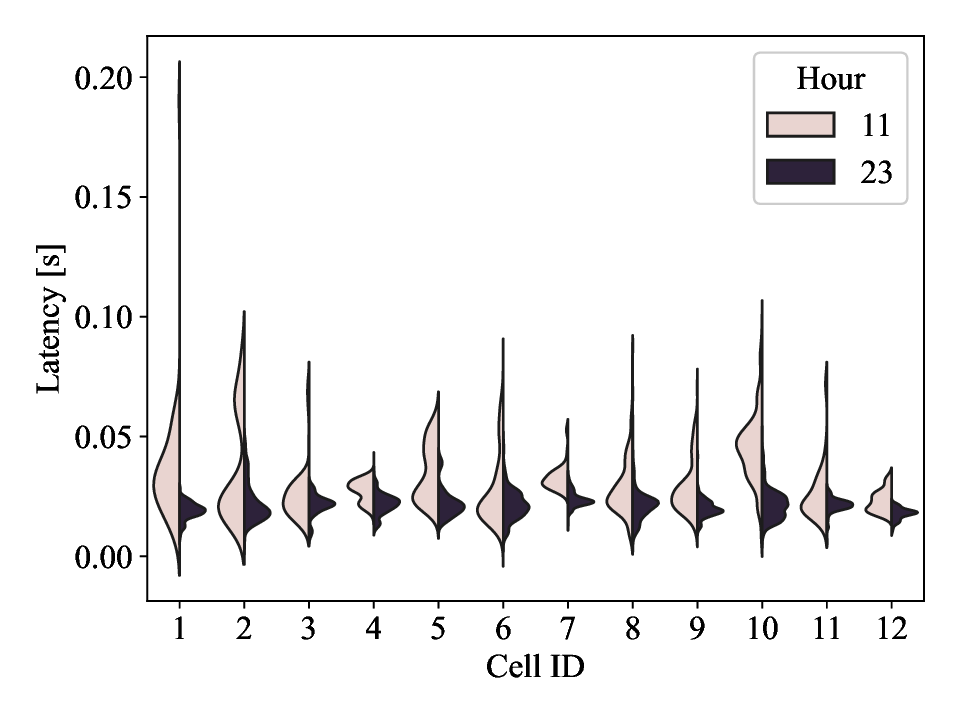}
    \caption{Violin plots showing the distribution of response latencies across individual cells at around 11:00 and 23:00. Each violin represents the latency distribution for a single cell, illustrating differences in latency times among cells.}
    \label{fig:latency_var}
\end{figure}

To examine the spatiotemporal heterogeneity of the generated dataset, we display the distribution of end-to-end latency across different cells at 11:00 and 23:00 in the violin plot in Fig.~\ref{fig:latency_var}. While several cells exhibit similar median latency levels, substantial differences are observed in both the spread and tail behavior of the distributions. This variability reflects the diversity among cell-specific conditions, such as traffic load and radio environment, which are intentionally modeled to evolve differently across time and space. The figure demonstrates that the proposed simulation framework generates varying latency distributions across cells at a fixed time instant, providing a suitable basis for studying spatially heterogeneous QoS.

\subsection{Predictive Experiment}
To assess the impact of spatial heterogeneity from the generated dataset~\cite{81qg-r354-25} on long-term latency prediction, we conduct a small-scale predictive experiment using the generated data. The objective of the experiment is to predict the parameters of the latency distribution in a given cell one hour ahead, based on historical observations.

We compare three predictive approaches. The first is a global approach, where a single predictive model is trained using the aggregated dataset collected from all cells. This model aims to capture common patterns shared across the network and is subsequently used to generate predictions for every cell. The second predictor is a local approach, where an independent predictive model is trained for each cell using only its locally generated data. In this case, the model parameters are trained to reflect the dynamics specific to each cell. The third approach serves as a na\"ive baseline, where the future distribution parameters are assumed to be identical to the current ones.

For the illustrative QoS prediction example, we implement a multi-layer perceptron with two hidden layers of 256 and 128 units, respectively, using ReLU activations and a 20\% dropout between layers to mitigate overfitting. The model is trained to predict the Gaussian distribution parameters of the latency, based on input features that include network-level, mobility-level, packet-level, and radio-level measurements. Training is performed using the Adam optimizer with a learning rate of $10^{-3}$ and a batch size of 32, minimizing the mean squared error over the dataset. The dataset is split into 80\% training dataset and 20\% test dataset. The dataset consists of approximately 90,000 samples collected over a 24-hour simulation with 1 s sampling granularity.

\begin{figure}
    \centering
    \includegraphics[width=0.65\linewidth]{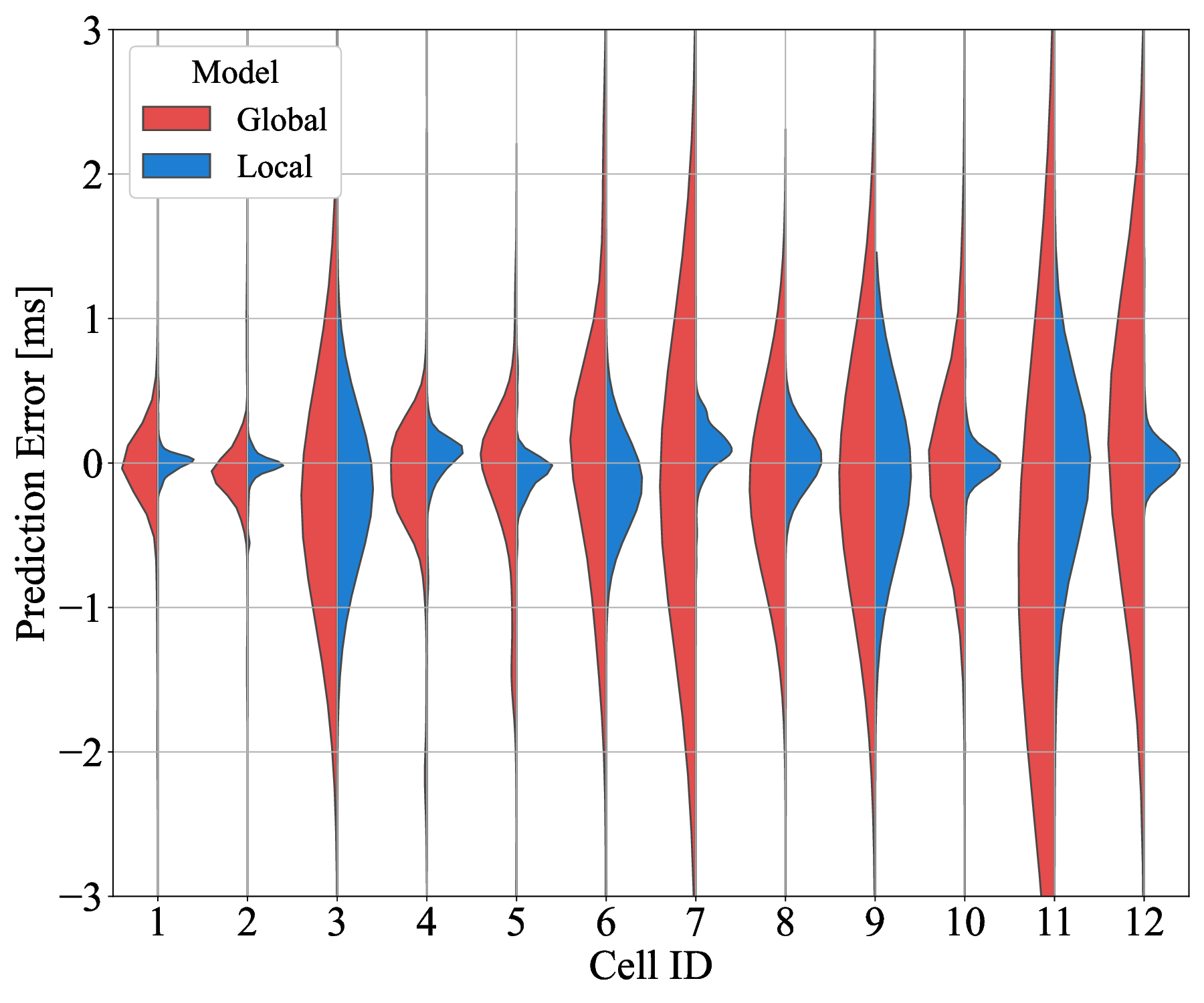}
    \caption{Violin plot of the prediction error across cells for the global and local predictive approaches. The local cell-level predictor achieves consistently lower errors and reduced variability compared to the single global model, indicating the presence of cell-specific dynamics and concept drift in the dataset.}
    \label{fig:pred_violin}
\end{figure}

The mean squared error of the predicted latency distributions is 0.61 ms for the naive baseline, 0.59 ms for the global model, and 0.16 ms for the local model, highlighting the substantial improvement achieved by the cell-specific approach. The prediction errors of the mean of the latency distributions from the global and local approaches are presented in the violin plot in Fig.~\ref{fig:pred_violin} and indicate that the local predictive approach consistently outperforms the global model across cells. For some cells, such as cells 7 and 12, the gain of the local approach is significant compared to the global approach. For other cells, such as cells 3 and 9, the gain is not as significant. However, this performance gap suggests that the latency dynamics exhibit significant cell-specific characteristics that are not fully captured by a single global model. In particular, the superior performance of the local approach points to the presence of concept drift across cells, where the underlying data differ spatially and evolve in distinct ways over time. 

These findings highlight the limitations of globally trained predictive models in heterogeneous wireless environments and point towards interesting directions for future studies using the dataset introduced in this work. Although the analysis is limited to the scenarios generated by the proposed NDT framework, the breadth of samples and the stability of the observed trends support the representativeness of the results within the considered urban vehicular setting. The violin plots aggregate prediction errors over a large number of samples collected across time and space, suggesting that the observed performance differences are statistically representative rather than driven by isolated events.  

\section{Discussion} 
\label{sec:dis}
The presented NDT simulator and the generated dataset constitute an important contribution to the research community because they provide a controlled, reproducible, and realistic environment for research in next-generation wireless networks. Unlike conventional NDTs that rely on stochastic channel models or limited field measurements, this work provides a rich and easily configurable framework that directly reflects the published dataset~\cite{81qg-r354-25}. The framework generates data that closely captures the dynamics of real urban deployments, while remaining user-friendly to adapt to desired scenarios and user requirements. This level of flexibility is essential for developing and benchmarking ML models that must operate reliably under varying radio conditions.

A broad range of studies on wireless network performance and data-driven network management can be enabled by the provided NDT framework. Because it captures information including mobility patterns, traffic conditions, and detailed radio-level measurements, it can be used to analyze how dynamic user behavior and environmental factors affect QoS in heterogeneous urban deployments.
Future work could benchmark anomaly detection algorithms to identify spatiotemporal patterns of congestion or assess the effectiveness of adaptive resource allocation strategies under sudden changes in the network. 
The controlled yet realistic NDT environment also enables researchers to evaluate the robustness of online learning algorithms under varying mobility and load conditions, thereby facilitating the development of algorithms that improve performance in next-generation wireless networks. 

This work presents an NDT framework to generate realistic wireless communication data. In this context, we define an NDT as a framework that generates spatially heterogeneous data with variations of the network load closely aligned with reality. We limit this work to building and providing the NDT framework, along with an example of a generated dataset from the NDT. A comparison of the provided NDT framework and the real world is inherently difficult at the level of fidelity targeted by this work. A meaningful validation would require access to synchronized measurements across radio, network, and application layers, together with detailed knowledge of base station configurations, scheduling policies, traffic demand, mobility traces, and the physical environment. In practice, these data are rarely publicly available due to privacy, commercial, and regulatory constraints. If partial measurements are available, they are often incomplete, noisy, or temporally misaligned across layers. 
Future work might be inclined to conduct a targeted measurement campaign in a confined area, where network parameters and mobility patterns can be tightly controlled, to conduct a quantitative comparison of the NDT framework to real-world data.

The computational requirements of the proposed NDT framework depend primarily on the number of network entities, the complexity of the ray-tracer, the sampling frequency, and the duration of the simulated scenario. While the integration of Sionna with ns-3 enables realistic propagation modeling, it increases computational load compared to purely stochastic models. In practice, simulation runtime scales roughly linearly with the number of vehicles $V$ and base stations $B$ as $\mathcal{O}(VB)$. However, urban areas may involve multiple rays due to reflections, diffractions, and scattering. More obstacles mean more rays to trace for each timestep, which adds extra computational cost per link. Memory usage similarly grows with scenario size, as channel matrices and node states must be stored over the sampling duration. 
We recommend GPU acceleration for running the ray-tracing simulations, which significantly helps to reduce runtime. While the framework is designed for accessibility and ease of use, researchers should balance simulation fidelity and scale to ensure feasible execution times, and potentially leverage parallelization or hybrid stochastic–deterministic channel models for large-scale experiments.

\section{Conclusion}
\label{sec:con}
We have presented a hybrid NDT framework that integrates ray-tracing with a discrete-event simulator to generate network datasets for vehicular communication. By combining site-specific channel modeling, dynamic network profiles, and heterogeneous mobility patterns, this system captures both spatial and temporal variations in network conditions across multiple layers of the network stack. The resulting NDT framework and dataset enable reproducible research and provide a rich testbed for training and evaluating ML models for QoS prediction, network optimization, and intelligent resource management. We provide suggestions for future studies based on this dataset and present an illustrative example that examines the role of predictive QoS model granularity for mitigating concept drift.

This work lowers the barrier to conducting controlled yet realistic experiments in urban and vehicular networks, offering a flexible platform for exploring heterogeneous traffic and varying deployment scenarios. We publish our simulator and dataset to promote reproducible, AI-driven research in next-generation communications systems.


\renewcommand{\bibname}{References}
\printbibliography

\end{document}